\documentclass[]{mn2e}
\usepackage{graphicx}
\usepackage{epsfig}

\newcommand \cm           {\,{\rm cm}}

\newcommand \g            {\,{\rm g}}
\newcommand \K            {\,{\rm K}}
\newcommand \pc           {\,{\rm pc}}

\newcommand \sr            {\,{\rm sr}}

\newcommand \mum          {\,{\rm \mu m}}

\newcommand \simali       {\sim\,}

\newcommand \NHw  {N_{\rm H}^w}
\newcommand \NHh  {N_{\rm H}^h}
\newcommand \Tw  {T_w}
\newcommand \Th  {T_h}
\newcommand \Cabsw  {C_{\rm abs}^w(a,\lambda)}
\newcommand \Cabsh  {C_{\rm abs}^h(a,\lambda)}
\newcommand \Ndw  {N_{\rm dust}^w}
\newcommand \Ndh  {N_{\rm dust}^h}
\newcommand \sgmabsw  {\sigma_{\rm abs}^w(\lambda)}
\newcommand \sgmabsh  {\sigma_{\rm abs}^h(\lambda)}
\newcommand \omegaw  {\Omega_w}
\newcommand \omegah  {\Omega_h}

\title[On the Anomalous Silicate Feature of NGC\,1068]
      {On the Anomalous Silicate Absorption Feature of 
       the Prototypical Seyfert 2 Galaxy NGC\,1068} 

\author[M.~K\"ohler \& A.~Li]
       {M.~K\"ohler and Aigen Li
       \thanks{%
                E-mail: {\sf koehlerme@missouri.edu,
                        lia@missouri.edu}
                }\\
      Department of Physics \& Astronomy, 
      University of Missouri, Columbia, MO 65211, USA
      }
\begin{document}
\date{Received date  / Accepted date }
\pagerange{\pageref{firstpage}--\pageref{lastpage}} \pubyear{2009}

\maketitle

\label{firstpage}
\begin{abstract}
The first detection of the silicate absorption
feature in AGNs was made at 9.7$\mum$ for
the prototypical Seyfert 2 galaxy NGC 1068 over
30 years ago, indicating the presence of a large column of
silicate dust in the line-of-sight to the nucleus.
It is now well recognized that type 2 AGNs exhibit 
prominent silicate absorption bands, 
while the silicate bands of type 1 AGNs 
appear in emission. More recently, using the Mid-Infrared 
Interferometric Instrument 
on the Very Large Telescope Interferometer,
Jaffe et al.\ (2004) by the first time spatially resolved
the parsec-sized dust torus around NGC 1068 and found that
the 10$\mum$ silicate absorption feature of the innermost
hot component exhibits an anomalous profile differing from
that of the interstellar medium and that of common olivine-type 
silicate dust. While they ascribed the anomalous absorption
profile to gehlenite (Ca$_2$Al$_2$SiO$_7$, a calcium aluminum 
silicate species), we propose a physical dust model and argue
that, although the presence of gehlenite is not ruled out, 
the anomalous absorption feature mainly arises from 
silicon carbide.
\end{abstract}


\section{Introduction}
%
%
Amorphous silicate is a major dust component of
the Galactic interstellar medium (ISM), as revealed 
directly by the smooth, featureless 9.7$\mum$ Si--O 
stretching and 18$\mum$ O--Si--O bending vibrational 
absorption bands, and indirectly by the depletion of
the silicate-forming elements Si, Mg, Fe and O in 
the gas phase (see Draine 2003). 
The first detection of the silicate absorption
feature in active galactic nuclei (AGNs) was made 
at 9.7$\mum$ for the prototypical Seyfert 2 galaxy 
NGC 1068 35 years ago by Rieke \& Low (1975)
and Kleinmann et al.\ (1976),
indicating the presence of a large column of
silicate dust in the line-of-sight to the nucleus.
It is known now that {\it most} of the type 2 AGNs 
(which only exhibit narrow emission lines in 
their optical spectra) display silicate 
absorption bands,
while the silicate features of type 1 AGNs 
(which are characterized by the presence of
both broad and narrow emission lines in their 
optical spectra) appear in emission.
This is consistent with the unified theory of AGNs which 
suggests that essentially all AGNs were ``born equal'': 
as illustrated in Figure 1a, all types of AGNs are surrounded 
by an optically thick dust torus and are basically the same 
object but viewed from different lines of sight
(see e.g. Antonucci 1993; Urry \& Padovani 1995).

\begin{figure*}
\begin{center}
\psfig{figure=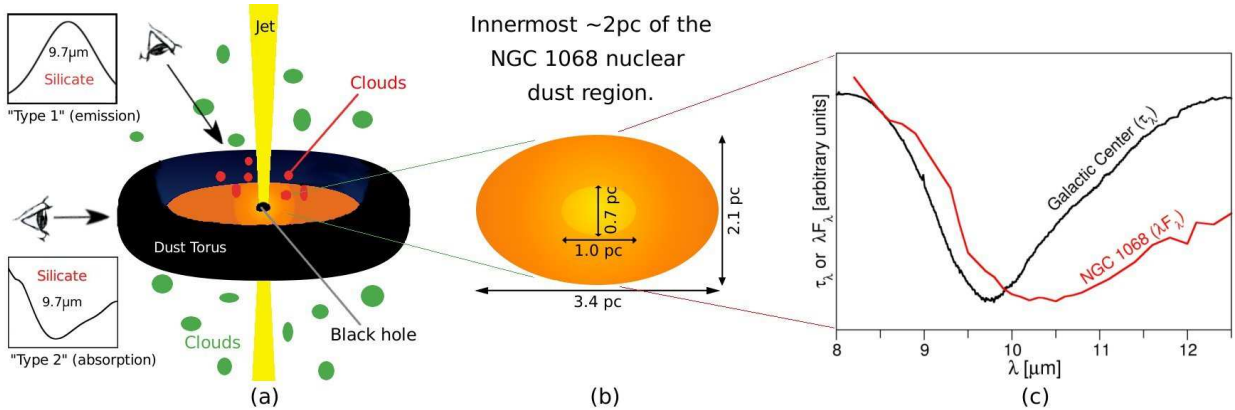, width=6.5in, angle=0}
\end{center}
\vspace{-3mm}
\caption{ \label{fig:cartoon}
         {\bf (a):} A schematic illustration of the dust torus
         around AGNs. The unified model of AGNs predicts 
         the silicate feature to appear in emission
         in type 1 AGNs (which are viewed face-on),
         and in absorption in type 2 AGNs (which are viewed edge-on).
         {\bf (b):} A schematic illustration of the innermost
         $\simali$2\,pc dust structure of the near nuclear region
         of the prototypical Seyfert galaxy NGC\,1068.
         The interferometric observations of Jaffe et al.\ (2004)
         resolved this dust structure into two components:
         a well-resolved 2.1\,pc$\times$3.4\,pc warm component
         and an inner hot component of $\simali$0.7\,pc along
         the jet and $<$\,1\,pc across.
         {\bf (c)}: A comparison of the Galactic center
         silicate absorption optical depth profile 
         (Kemper et al.\ 2004) with the interferometric 
         mid-IR spectrum of the innermost $\simali$2\,pc 
         near nuclear dust region of NGC\,1068 
         (Jaffe et al.\ 2004).
         }
\end{figure*}

Within the framework of the unified theory of AGNs,
for type 1 AGNs which are viewed face-on, one would expect to see 
the silicate features in {\it emission} since the silicate dust 
in the surface of the inner torus wall will be heated to 
temperatures of several hundred kelvin to $\simali$1000\,K
by the radiation from the central engine, allowing for 
a direct detection of the 9.7$\mum$ and 18$\mum$ silicate 
bands emitted from the dust (see Figure 1a). 
Their detection has been reported 
with {\it Spitzer} (Hao et al.\ 2005; 
Siebenmorgen et al.\ 2005; Sturm et al.\ 2005; 
Weedman et al.\ 2005; Shi et al.\ 2006; Schweitzer et al.\ 2008). 
For type 2 AGNs which are viewed edge-on,
the silicate features are mostly seen in {\it absorption}
(see Figure 1a) because of the obscuration of the optically 
thick dust torus (e.g. see Roche et al.\ 1991, 2007; 
Siebenmorgen et al.\ 2004; Hao et al.\ 2007; Spoon et al.\ 2007).

We note that the silicate emission features have 
     also been detected in several type 2 QSOs 
     (Sturm et al.\ 2006, Teplitz et al.\ 2006).
     Efstathiou (2006), Marshall et al.\ (2007)
     and Schweitzer et al.\ (2008) suggested that
     the silicate emission may actually arise from
     the narrow-line region (NLR) dust.
However, Nikutta et al.\ (2009) argued that
     the clumpy dust torus model could explain
     the detection of the silicate feature in emission
     in type 2 sources. 
More recently, by combining {\it Gemini} and {\it Spitzer} 
     mid-IR imaging and spectroscopy of NGC\,2110 (the closest 
     known Seyfert 2 galaxy with silicate emission features), 
Mason et al.\ (2009) were able to constrain the location of 
the silicate-emitting region to within 32\,pc of the nucleus. 
     Their results were consistent with both a NLR origin 
     and an edge-on torus origin of the silicate emission.

The silicate emission profiles of AGNs are rather diverse,
suggesting that the AGN silicate grains were probably not
``born equal''.
Some AGNs display an ``anomalous'' Si--O stretching band
which is considerably shifted to longer wavelengths
(from the canonical 9.7$\mum$ to $\simali$11.5$\mum$;
hereafter we call it ``redshift'') and broadened 
in comparison with the Galactic silicate profile 
(Hao et al.\ 2005, Siebenmorgen et al.\ 2005, 
Sturm et al.\ 2005, Smith et al.\ 2010),
with the degree of redshifting and broadening 
varying from one AGN to another.

Some AGNs, e.g. 3C\,273 (Hao et al.\ 2005) 
and NGC\,7213 (Wu et al.\ 2009), exhibit an unusually strong 
red tail of the 18$\mum$ O--Si--O bending band, 
incomparable with the Galactic silicate profiles.
The silicate {\it absorption} profiles of some AGNs also 
show appreciable deviations from that of the Galactic ISM.
More specifically, using the {\it Mid-Infrared Interferometric 
Instrument} (MIDI) on the ESO's {\it Very Large Telescope 
Interferometer} (VLTI), 
Jaffe et al.\ (2004) by the first time spatially resolved 
the parsec-sized dust torus around NGC\,1068 (see Figure 1b)
and found that the 10$\mum$ silicate absorption feature of 
the innermost hot component exhibits an anomalous profile 
differing from that of the ISM (see Figure 1c): 
the 9.7$\mum$ silicate absorption spectrum of NGC\,1068 
shows a relatively flat profile from 8 to 9$\mum$ 
and then a sharp drop between 9 and 10$\mum$, 
while the Galactic silicate absorption profiles
begin to drop already at about 8$\mum$.
Jaffe et al.\ (2004) found that the profile of the silicate 
absorption towards the hot component does not fit well to 
the profiles of common olivine-type silicate dust
(also see Raban et al.\ 2009).
They obtained a much better fit using the profile of 
gehlenite (Ca$_2$Al$_2$SiO$_7$), 
a calcium aluminum silicate species
(a high-temperature dust species found
in some supergiant stars, e.g. see Speck et al.\ 2000).

As a part of a systematic study of the nature of 
the dust in AGNs and particularly of the AGN dust
mineralogy, we quantitatively model the anomalous
silicate absorption feature associated with the innermost 
hot component of the dust torus of NGC\,1068.


\section{NGC\,1068\label{sec:background}}
At a distance of only 14.4\,Mpc, 
NGC\,1068 is one of the closest and probably the
most intensely studied Seyfert 2 galaxy.
It has played a key role in the establishment
of the unified model of AGNs: classified as a Seyfert 2 
based on the presence of narrow emission lines and absence 
of broad emission lines in its optical spectrum,
NGC\,1068 is known to also harbor an obscured 
Seyfert 1 nucleus as revealed by the detection 
of broad emission lines in polarized light 
(Antonucci \& Miller 1985).
As expected from an edge-on geometry of the proposed 
dust torus for type 2 AGNs, NGC\,1068 has long been
known to display a strong absorption band at 9.7$\mum$,
attributed to amorphous silicate dust
(Rieke \& Low 1975, Kleinmann et al.\ 1976, 
Roche et al.\ 1991, Lutz et al.\ 2000,
Le Floc'h et al.\ 2001, Siebenmorgen et al.\ 2004).

Jaffe et al.\ (2004) obtained the 8--13.5$\mum$ $N$-band 
spectra of the nuclear region of NGC\,1068, using the MIDI
instrument coupled to the VLTI [with a spatial resolution of 
$\simali$10\,milliarcsec (mas) at $\lambda$\,=\,10$\mum$, 
corresponding to $\simali$0.7\,pc]. 
Their interferometric mid-IR observations by the first time 
spatially resolved the central parsec-sized circumnuclear 
dust structure of NGC\,1068 (see Figure 1b).
%


Rhee \& Larkin (2006) obtained spatially-resolved 
mid-IR spectra of the nucleus of NGC\,1068, using 
the {\it Long Wavelength Spectrograph} (7.71--12.48$\mum$) 
of the Keck I 10\,m telescope 
(with a spatial resolution of 0.25\arcsec, 
corresponding to $\simali$18\,pc in physical scale).
They found that the silicate absorption feature varies 
over the nuclear region: the depth of the feature reaches 
its maximum in the nucleus and decreases with distance 
from the central engine.
Spatially-resolved mid-IR spectra of the nucleus of 
NGC\,1068 were also obtained by Mason et al.\ (2006),
using the {\it Michelle} $N$-band (7--14$\mum$) 
spectrometer of the Gemini North 8.1\,m telescope
(with a spatial resolution of 0.4\arcsec, 
corresponding to $\simali$30\,pc).
They also found that the silicate feature profile 
and depth exhibit striking spatial variations.



The interferometric observations of Jaffe et al.\ (2004) 
resolved the near-nuclear mid-IR emission into 
a {\it warm} component ($\simali$320\,K) in 
a (well-resolved) structure $\simali$2.1\,pc thick 
and $\simali$3.4\,pc in diameter, surrounding 
a smaller (marginally-resolved) hot structure ($>800\K$)
of $\simali$0.7\,pc in size along the jet and $<1\pc$ across
(also see Raban et al.\ 2009). 
%
%
They found that the silicate absorption feature could not
be fit with the Galactic silicate profile or the profiles 
of common olivine-type silicate dust (see Figure 1c), 
as they begin to drop already at $\simali$8$\mum$, 
while the interferometric spectrum of NGC\,1068 
(with the highest spatial resolution) 
shows a relatively flat profile from 8 to 9$\mum$
(and then a sharp drop between 9 and 10$\mum$). 
Therefore, they concluded that {\it their observations 
would require special dust properties in the innermost 2\,pc 
around the nucleus of NGC\,1068.} 
%
They obtained a quite satisfactory fit with the profile
of gehlenite Ca$_2$Al$_2$SiO$_7$, 
a high-temperature dust species.
However, the approach taken by Jaffe et al.\ (2004)
seems to us too simplified (see \S5).  
This stimulates us to present a more physical model 
(see \S3) to interpret the unusual silicate absorption 
spectrum of NGC\,1068 obtained by Jaffe et al.\ (2004).

\begin{figure}
\begin{center}
\psfig{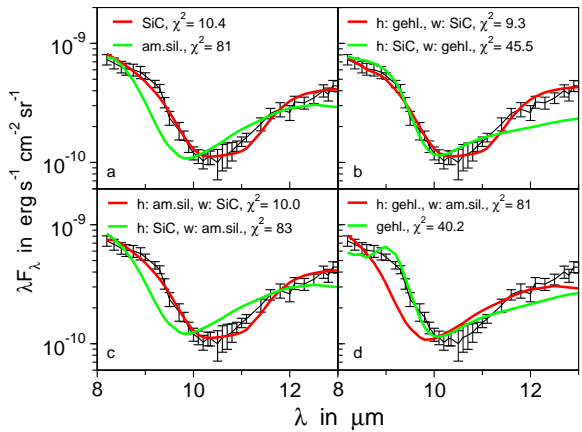}
\end{center}
\vspace{-3mm}
\caption{\label{fig:model1}
         Comparison of the interferometric spectrum
         of the innermost, parsec-sized dusty nuclear 
         region of NGC\,1068 (Jaffe et al.\ 2004) with 
         the model spectra calculated from various dust species:
         (a) SiC (red line)
             or amorphous olivine silicate dust (green line)
             for both the inner hot component 
             and the surrounding warm component;
         (b) gehlenite in the hot component 
             and SiC in the warm component (red line),
             or SiC in the hot component and gehlenite
             in the warm component (green line);
         (c) amorphous olivine in the hot component 
             and SiC in the warm component (red line),
             or SiC in the hot component and amorphous olivine
             in the warm component (green line); and
         (d) gehlenite in the hot component 
             and amorphous olivine in the warm component (red line),
             or gehlenite in both the hot and warm components
             (green line).
         } 
\end{figure}

\vspace{-5mm}
\section{The Model\label{sec:model}}
Following Jaffe et al.\ (2004), we assume that the nuclear 
dust region of NGC\,1068 consists of two components:
a hot component and a warm component.
Let $\NHw$ and $\NHh$ be the hydrogen column densities of
the warm and hot components, respectively;
$\Tw$ and $\Th$ be the (mean) dust temperatures of
the warm and hot components, respectively;
$\Cabsw$ and $\Cabsh$ be the absorption cross sections
of the dust species of size $a$ at wavelength $\lambda$
of the warm and hot components, respectively.
For simplicity, we will assume a single grains size
of $a=0.1\mum$ for both the warm and hot components.
Let $\Ndw$ and $\Ndh$ be the column densities of the dust 
(of size $a$) of the warm and hot components, respectively. 
Let $\sgmabsw$ and $\sgmabsh$ be the {\it total} 
absorption cross sections per H nucleon of the dust at wavelength 
$\lambda$ of the warm and hot components, respectively.
Apparently, we have $\sgmabsw = \Cabsw\times\left(\Ndw/\NHw\right)$
and $\sgmabsh = \Cabsh\times\left(\Ndh/\NHh\right)$.
Let $\omegaw$ and $\omegah$ be the solid angles extended 
by the warm and hot components, respectively.
The flux emitted by the warm and hot components
received at the Earth is given by\footnote{%
  Consider a dusty system with a hydrogen column density 
  of $N_{\rm H}$.
  Let $\sigma_{\rm abs}(\lambda)$ be the total dust absorption
  cross section per H nucleon.
  Let $p_\lambda$ be the power radiated per solid angle
  in $\left[\lambda, \lambda+d\lambda\right]$ per H.
  With the dust self-absorption taken into account,
  the emission intensity is
  $I_\lambda = \int_{0}^{N_{\rm H}}
   p_\lambda \exp\left[-N_{\rm H}^{\prime}
   \sigma_{\rm abs}(\lambda)\right]
   dN_{\rm H}^{\prime}
   = p_\lambda \left\{1-\exp\left[-N_{\rm H}
     \sigma_{\rm abs}(\lambda)\right]\right\}
     /\sigma_{\rm abs}(\lambda)$.
   If the dusty system is optically thin
   [i.e. $N_{\rm H} \sigma_{\rm abs}(\lambda)\ll 1$],
   $I_\lambda\approx p_\lambda N_{\rm H}$.
   }
%
\begin{eqnarray}
\nonumber
F_\lambda & = & \frac{\omegaw}{4\pi}
B_{\lambda}(\Tw) 
\left\{1-\exp\left[-\NHw\sgmabsw\right]\right\}\\
\nonumber
&& + \frac{\omegah}{4\pi}
B_{\lambda}(\Th) 
\left\{1-\exp\left[-\NHh\sgmabsh\right]\right\}\\
&&\times \exp\left[-\NHw\sgmabsw\right] ~~.
\end{eqnarray}
The $\exp\left[-\NHw\sgmabsw\right]$ term in eq.2
accounts for the fact that the emission from
the inner hot component is further absorbed by
the dust in the warm component (like a screen).
The solid angles of the warm and hot components are
taken to be
$\omegaw = 30\,{\rm mas} \times 49\,{\rm mas}
\approx 3.45\times 10^{-14}\sr$,
and
$\omegah = 10\,{\rm mas} \times 12\,{\rm mas}
\approx 2.82\times 10^{-15}\sr$ 
(Jaffe et al.\ 2004).

We take the column density of the dust of a given species 
to be limited by the cosmic abundance of the rarest element 
contained in a given species (e.g. Ca in gehlenite Ca$_2$Al$_2$SiO$_7$). 
Therefore, the total number of gehlenite grains
(per H) can not exceed
\begin{equation}
\frac{N_{\rm geh}}{N_{\rm H}}
= \frac{1}{2}
  \frac{\left[{\rm Ca}/{\rm H}\right]}
  {\left(4\pi/3\right)a^3\rho_{\rm geh}/
  \left(\mu_{\rm geh} m_{\rm H}\right)} ~,
\end{equation}
where $\left[{\rm Ca}/{\rm H}\right]$ is the Ca abundance 
(relative to H), $\rho_{\rm geh} =2.91\g\cm^{-3}$ is 
the mass density of gehlenite (Mutschke et al.\ 1998), 
$\mu_{\rm geh} = 274$ is the molecular weight 
of gehlenite (Ca$_2$Al$_2$SiO$_7$), and
$m_{\rm H}$ is the atomic mass of H.
Similarly, for amorphous olivine-type 
(MgFeSiO$_4$) silicate dust,
the total numbers of silicate and SiC grains (per H)
are limited to
\begin{equation}
\frac{N_{\rm sil}}{N_{\rm H}}
 = \frac{\left[{\rm Si}/{\rm H}\right]}
  {\left(4\pi/3\right)a^3\rho_{\rm sil}/
  \left(\mu_{\rm sil} m_{\rm H}\right)} ~,~
\frac{N_{\rm sic}}{N_{\rm H}}
 = \frac{\left[{\rm Si}/{\rm H}\right]}
  {\left(4\pi/3\right)a^3\rho_{\rm sic}/
  \left(\mu_{\rm sic} m_{\rm H}\right)} ~,
\end{equation}
where $\left[{\rm Si}/{\rm H}\right]$ is 
the Si abundance (relative to H),
$\rho_{\rm sil} =3.5\g\cm^{-3}$ is the mass density of 
MgFeSiO$_4$, $\mu_{\rm sil} = 172$ is the molecular 
weight of MgFeSiO$_4$,
$\rho_{\rm sic} =3.2\g\cm^{-3}$ is the mass density of 
silicon carbide dust, and $\mu_{\rm sic} = 40$ is 
the molecular weight of SiC.
%

In modeling the interferometric mid-IR absorption spectrum, 
we first need to specify the dust compositions for the warm
and hot components. Once the dust compositions are specified,
we obtain their absorption cross sections from Mie theory
(Bohren \& Huffman 1983). We are then left with only four
parameters: $T_w$, $T_h$, $N_{\rm H}^w$, and $N_{\rm H}^h$.  

\begin{figure}
\begin{center}
\psfig{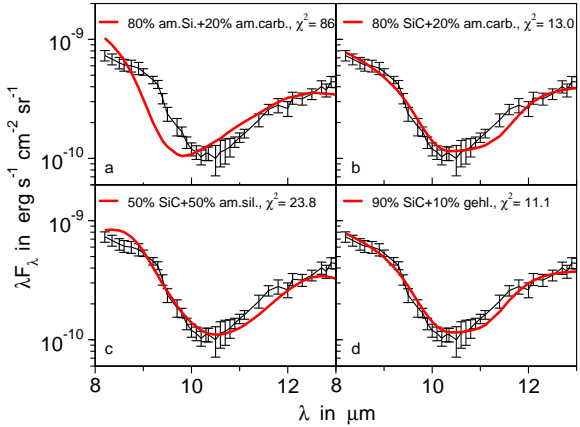}
\end{center}
\vspace{-3mm}
\caption{
         \label{fig:model2}
         Comparison of the interferometric spectrum
         of the innermost, parsec-sized dusty nuclear 
         region of NGC\,1068 with the model spectra 
         calculated from
         (a) a mixture of 80\% amorphous olivine silicate dust 
             and 20\% amorphous carbon;
         (b) a mixture of 80\% SiC 
             and 20\% amorphous carbon; 
         (c) a mixture of 50\% SiC and 50\% amorphous olivine; and
         (d) a mixture of 90\% SiC
             and 10\% gehlenite.
         We assume the same dust composition for
         both the hot and warm regions. 
         }
\end{figure}

\vspace{-5mm}
\section{Results\label{sec:results}}
We have considered a wide variety of dust materials,
including amorphous olivine (Dorschner et al.\ 1995),
amorphous pyroxene (J\"ager et al.\ 1994),
amorphous carbon (Rouleau \& Martin 1991), 
$\alpha$-SiC (Laor \& Draine 1993), 
glassy gehlenite (Mutschke et al.\ 1998), 
and spinel (Fabian et al.\ 2001).
We have also considered mixtures of 
these materials, with their absorption properties
calculated from Mie theory in combination with 
the Bruggeman effective medium theory (Bohren \& Huffman 1983). 
For the abundance constraints (see eqs.3,4),
we adopt the solar abundance of
$\left[{\rm Ca/H}\right]\approx 2.2 \cdot 10^{-6}$ 
and $\left[{\rm Si/H}\right]\approx 31.6 \cdot 10^{-6}$
(Asplund et al.\ 2009). 
With $a$\,=\,0.1$\mum$,
$N_{\rm gel}/N_{\rm H}\approx 4.1\cdot 10^{-14}$,
$N_{\rm sil}/N_{\rm H}\approx 6.2\cdot 10^{-13}$, and
$N_{\rm sic}/N_{\rm H}\approx 1.6\cdot 10^{-13}$
(see eqs.2,3).


As shown in Figure 2, the best fits to the interferometric mid-IR 
spectrum of Jaffe et al.\ (2004) are provided by models which all 
require SiC (see Table 1 for the model parameters): 
either with pure SiC in both the hot region and the surrounding
warm region (see Figure 2a), or with gehlenite in the hot region and
SiC in the warm region (see Figure 2b), or with amorphous olivine
in the hot region and SiC in the warm region (see Figure 2c). 
The composition of the inner hot component is less well constrained 
as the large column density required for the inner component
causes the dust there to emit effectively like a blackbody.   
The absorption profile is predominantly provided by the SiC dust 
in the outer warm region. 

We have also tried pure amorphous olivine dust for both regions. 
As shown in Figure 2a, even the best fit fails to fit the observed
spectrum, particularly in the blue wing. 
With the warm region filled with amorphous olivine
and with the hot region filled with either SiC (see Figure 2c)
or gehlenite (see Figure 2d), we obtain model spectra similar
to that of the pure amorphous olivine model, confirming that
it is the dust in the outer warm region dominates 
the absorption profile.

We have also considered gehlenite. Assuming either gehlenite
for both regions (see Figure 2d) or gehlenite only for the outer
warm region and other dust species (e.g. SiC) for the inner hot 
region (see Figure 2b), we see that good fits to the blue wing 
of the 9.7$\mum$ feature are achieved, but both models fail at
$\lambda>11\mum$.


Finally, we model the observed mid-IR spectrum with a mixture
of two dust species and assume the same mixture for 
both the hot and warm regions (see Figure 3). 
For the amorphous olivine-amorphous carbon mixture, 
even the closest fit (given by a mixture of silicate
of a volume fraction of 80\% and amorphous carbon 
of a volume fraction of 20\%) fails to reproduce 
the blue wing of the 9.7$\mum$ feature (see Figure 3a).
Better fits could be achieved if the mixture contains SiC
(see Figure 3b,c,d).  

\begin{table}
{\scriptsize
\begin{center}
\begin{tabular*}{0.53\textwidth}{lcccccl}
\hline\hline
Hot & Warm   
& $N_{\rm H}^h$ & $N_{\rm H}^w$ & $T_h$  &  $T_w$ & $\chi^2$\\
Dust & Dust    
& [$\cm^{-2}$] & [$\cm^{-2}$] & [K]  &  [K] &  \\
\hline
{\bf SiC} & {\bf SiC}  &  ${\bf 4.0 \cdot 10^{25}}$       & ${\bf 6.3 \cdot 10^{23}}$       & {\bf 1110  }     & {\bf 250 }       & {\bf 10.4} \\
sil. & sil.  & $2.5 \cdot 10^{24}$       & $7.9 \cdot 10^{22}$       & 870        & 100        & 81 \\
{\bf gehl.} & {\bf SiC}        & ${\bf 1.6 \cdot 10^{25}}$       & ${\bf 6.3 \cdot 10^{23}}$       & {\bf 1140}       & {\bf 250}        & {\bf 9.3}  \\
SiC & gehl.        & $4.0 \cdot 10^{25}$       & $6.3 \cdot 10^{23}$       & 840        & 220        & 45.5  \\
{\bf sil.} & {\bf SiC}      & ${\bf 2.5 \cdot 10^{24}}$       & ${\bf 6.3 \cdot 10^{23}}$       & {\bf 1110}       & {\bf 250}        & {\bf 10.0}  \\
SiC & sil.      & $4.0 \cdot 10^{25}$       & $7.9 \cdot 10^{22}$       & 870        & 100        & 83  \\
gehl. & sil.    & $2.5 \cdot 10^{25}$       & $7.9 \cdot 10^{22}$       & 870        & 100        & 81  \\
gehl. & gehl.   & $6.3 \cdot 10^{24}$       & $7.9 \cdot 10^{23}$       & 960        & 230        & 40.2 \\
80\% sil. & 20\% am.carb. & $1.6 \cdot 10^{25}$      & $1.3 \cdot 10^{23}$       & 1020        & 100        & 86  \\
80\% SiC & 20\% am.carb.     & $4.0 \cdot 10^{25}$      & $6.3 \cdot 10^{23}$       & 1080        & 250        & 13.0  \\
50\% SiC & 50\% sil.    & $1.6 \cdot 10^{23}$      & $1.6 \cdot 10^{23}$       & 1770        & 240        & 23.8  \\
90\% SiC &  10\% gehl.      & $6.3 \cdot 10^{25}$      & $1.0 \cdot 10^{24}$       & 1020        & 250        & 11.1 \\
\hline
\end{tabular*}
\caption{Model Parameters. The dust column densities $N_{\rm dust}$ 
         can be obtained from $N_{\rm H}$ using eqs.2,3
         (e.g. $N_{\rm gel} = N_{\rm H} \times 
               \left[N_{\rm gel}/N_{\rm H}\right]$).
         The dust masses $m_{\rm dust}$ can be derived from
         $N_{\rm dust}$ and the solid angle $\Omega$ 
         (e.g. $m_{\rm gel}$\,=\,0.5\,$\left[{\rm Ca/H}\right] 
               \mu_{\rm gel} m_{\rm H} N_{\rm H} \times \Omega d^2$).
         }   
\label{tab:data}
\end{center}
}
\end{table}

\vspace{-6mm}
\section{Discussion}\label{sec:discussion}
In \S4 we have shown that, for a physically plausible model,
to achieve any reasonably close fits to the interferometric 
spectrum of NGC\,1068, SiC is required. SiC is a highly 
refractory dust species. It has been identified in the outflows
of carbon stars. Interstellar SiC grains have been found intact 
in ``primitive '' meteorites (e.g. Bernatowicz et al.\ 1987),
although the fraction of silicon in the local diffuse ISM 
in SiC dust is small: $<$\,5\% (Whittet et al.\ 1990)
or $\simali$3\% (Min et al.\ 2007).
SiC was one of the three major dust species (i.e. amorphous 
silicate, graphite, and SiC) considered by Laor \& Draine (1993) 
in their extensive AGN dust modeling.
Although we are not able to rule out gehlenite as 
the major dust species in the inner hot component, 
SiC should be the dominant dust species in the outer warm 
component and dominates the mid-IR absorption. 

For all models, the determined temperatures of the hot and
warm components are $>800\K$ and $<300\K$, respectively
(see Table 1). For the SiC-containing model, the temperatures 
of the dust in the hot and warm components are $T_h \sim 1100\K$
and $T_w\sim 250\K$, respectively. 
The derived H column density of the hot component 
$\NHh$ is in the order of $\simali$10$^{25}\cm^{-2}$,
consistent with the measurements of the inner wall
by Matt et al.\ (1997, 2004). 
The warm component requires $\NHw$ to be 
in the order of $\simali$10$^{23}\cm^{-2}$,
consistent with the measurements of
Kinkhabwala et al.\ (2002).  

Finally, we note that although the calculations were 
carried out for a grain radius of 0.1$\mum$, 
increasing the grain radius or assuming a porous
structure causes the deviation between the observed
and model spectra to increase.

Jaffe et al.\ (2004) argued that a quite satisfactory fit 
to the interferometric mid-IR spectrum of the near-nuclear 
dust region of NGC\,1068 could be obtained with the absorption
profile of gehlenite.
Gehlenite is a silicate of the melilite group 
with a high condensation temperatures 
($\simali$1735\,K for melilite, see Hutchison 2004). 
It has been found in objects in the solar system: 
calcium-aluminum rich inclusions in carbonaceous 
chondrites (see e.g. Morlok et al.\ 2008), 
anhydrous interplanetary dust particles (Greshake et al.\ 1996), 
and cometary dust collected by the {\it Stardust} spacecraft
from comet Wild 2 (Zolensky et al.\ 2006, Schmitz et al.\ 2008). 
It is assumed that gehlenite was produced close to 
the young Sun at high temperatures and was distributed 
in the entire system.

In their model, Jaffe et al.\ (2004) assumed that 
(1) both the warm and hot components emit like a black-body, and
(2) each component is subject to an attenuation of
$\exp(-\tau_\lambda)$, with the optical depth profile
$\tau_\lambda$ varying to match that of known silicate species. 
Their modeling theme can be described as the following
\vspace{-1mm}
\begin{equation}
F_\lambda = c_w B_\lambda(T_w)\exp(-\tau_\lambda^w)
+ c_h B_\lambda(T_h)\exp(-\tau_\lambda^h)\exp(-\tau_\lambda^w) ~,
\end{equation}
\vspace{-1mm}
where $c_w$ and $c_h$ are the constants (related to 
the total dust masses) for the warm and hot 
components, respectively;\footnote{%
  More explicitly, under the assumption that the dust emits
  like a blackbody,
  $c_w = 3 m_w/\left(4\pi\rho_w d^2\right)$
  and $c_h = 3 m_h/\left(4\pi\rho_h d^2\right)$,
  where $m_w$ and $m_h$ are respectively the total dust masses
  of the warm and hot components, $\rho_w$ and $\rho_h$ are 
  respectively the mass densities of the dust in the warm and 
  hot components.
  }
$T_w$ and $T_h$ are the dust temperatures of 
the warm and hot components, respectively;
$B_\lambda(T)$ is the Planck function of temperature $T$
at wavelength $\lambda$;
$\tau_\lambda^{w}$ and $\tau_\lambda^{h}$ 
are the optical depths in front of
the warm and hot components, respectively.
Since the hot component is embedded in 
the warm component, the hot component is subject
to an attenuation of $\exp(-\tau_\lambda^h-\tau_\lambda^w)$.

While it was rather successful in fitting the MIDI spectra, 
the assumptions made by Jaffe et al.\ (2004) were
not justified: 
(1) their modeling theme implied that as if 
there were an absorbing screen in front of the hot component 
and an additional absorbing screen in front of the warm component; and
(2) if the grains in the hot and warm components indeed emit
like a blackbody as they assumed, the grains must be much 
larger than several micrometers and reach the geometrical 
optics regime.\footnote{%
  This requires the grain size $a$ satisfying
  the criterion of $2\pi a/\lambda \gg 1$ at 
  the wavelength range of the silicate Si--O band,
  i.e. $\lambda$\,$\sim$\,10$\mum$ (see Bohren \& Huffman 1983).
  } 
These grains would not produce any noticeable spectral
features around $\lambda=10\mum$, and therefore the silicate
absorption must arise from the absorbing screens.

To summarize, we agree with Jaffe et al.\ (2004) that 
the interferometric spectrum of NGC\,1068 cannot be 
explained with amorphous silicate with an olivine 
normative composition. However, unlike Jaffe et al.\ (2004) 
who attributed the unusual 9.7$\mum$ absorption 
feature of NGC\,1068 to gehlenite, we argue that
SiC is mostly responsible for this feature 
although the presence of gehlenite 
in the inner hot region is not ruled out.

\vspace{-3mm}
\section*{Acknowledgments}
We thank C. Kemper for providing us with
the ISO spectrum of the Galactic center,
and J.Y. Hu, M.P. Li, H.A. Smith, A.K. Speck, 
and the anonymous referee for very helpful 
discussions/comments. This work is supported 
in part by NASA/HST Theory Programs, 
NASA/Spitzer Theory Programs,
and a NASA/Chandra Theory program.

\bsp
\label{lastpage}

\end{document}